\input psfig.sty
\nopagenumbers 
\magnification=\magstep1 
\hsize 6.0 true in 
\hoffset 0.25 true in 
\emergencystretch=0.6 in 
\vfuzz 0.4 in 
\hfuzz 0.4 in 
\vglue 0.1true in 
\mathsurround=2pt 
\def\nl{\noindent} 
\def\nll{\hfil\break\noindent} 
\def\np{\hfil\vfil\break} 
\def\ppl#1{{\leftskip=9cm\noindent #1\smallskip}} 
\def\title#1{\bigskip\goodbreak\noindent\bf #1 ~ \trr\smallskip} 
 
\font\trr=cmr10 
\font\bf=cmbx10 
\font\bmf=cmmib10 
\font\sl=cmsl10 
\font\it=cmti10 
\font\trbig=cmbx10 scaled 1500 
\font\tiny=cmr8 
\def\ma#1{\hbox{\vrule #1}} 
\def\mb#1{\hbox{\bmf#1}} 
\def\ng{{>\kern -8pt|\kern 8pt}} 
\def\nle{{<\kern -8pt|\kern 8pt}} 
\def\hi#1#2{$#1$\kern -2pt-#2} 
\def\hy#1#2{#1-\kern -2pt$#2$} 

\def\sgn{{\rm sgn}}

\def\Ai{{\rm Ai}}
\def\rtitle{Relativistic N-boson systems} 
\def\ptitle{Relativistic N-boson systems bound by } 
\def\ptitlee{pair potentials $V(r_{ij}) = g(r_{ij}^2)$} 

\output={\shipout\vbox{\makeheadline\ifnum\the\pageno>1 {\hrule} \fi 
{\pagebody}\makefootline}\advancepageno} 
 
\headline{\noindent {\ifnum\the\pageno>1 
{\tiny \rtitle\hfil page~\the\pageno}\fi}} 
\footline{} 
\newcount\zz \zz=0 
\newcount\q 
\newcount\qq \qq=0 
 
\def\pref #1#2#3#4#5{\frenchspacing \global \advance \q by 1 
\edef#1{\the\q}{\ifnum \zz=1 { %
\item{[\the\q]}{#2} {\bf #3},{ #4.}{~#5}\medskip} \fi}} 
 
\def\bref #1#2#3#4#5{\frenchspacing \global \advance \q by 1 
\edef#1{\the\q}{\ifnum \zz=1 { %
\item{[\the\q]}{#2}, {\it #3} {(#4).}{~#5}\medskip} \fi}} 
 
\def\gref #1#2{\frenchspacing \global \advance \q by 1 
\edef#1{\the\q}{\ifnum \zz=1 { %
\item{[\the\q]}{#2}\medskip} \fi}}

\def\sref #1{~[#1]} 
 
\def\references#1{\zz=#1 
\parskip=2pt plus 1pt 
{\ifnum \zz=1 {\noindent \bf References \medskip} \fi} \q=\qq 
\pref{\bse}{E.~E.~Salpeter and H.~A.~Bethe, Phys.~Rev.}{84}{1232 (1951)}{} 
\pref{\se}{E.~E.~Salpeter, Phys.~Rev.}{87}{328 (1952)}{} 
\bref{\lieb}{E.~H.~Lieb and M.~Loss}{Analysis}{American Mathematical Society, 
New York, 1996} {The definition of the Salpeter kinetic-energy operator is 
given on p.~168;~Jensen's inequality is given in Theorem~2.2  on p.~38.} 
\pref{\hallpost}{R. L. Hall and H. R. Post, Proc. Phys. Soc. (Lond.)}{90}{381 (1967)}{}
\pref{\hallferm}{R. L. Hall, Proc. Phys. Soc. (Lond.)}{91}{16 (1967)}{}
\pref{\halla}{R.~L.~Hall, J.~Math.~Phys.}{24}{324 (1983)}{}
\pref{\hallb}{R.~L.~Hall, J.~Math.~Phys.}{25}{2708 (1984)}{}
\pref{\hallc}{R.~L.~Hall, W.~Lucha, and F.~F.~Sch\"oberl, J.~Math.~Phys.}
{42}{5228 (2001)}{} 
\pref{\halld}{R. L. Hall, W. Lucha, and F. F. Sch\"oberl, Int. J. Mod. Phys. A}{17}{1931 (2002)}{} 
\pref{\halle}{R. L. Hall, W. Lucha, and F. F. Sch\"oberl, 
J. Math. Phys.}{43}{5913 (2002)}{} 
\gref{\hallf}{R. L. Hall, W. Lucha, and F. F. Sch\"oberl, J. Math. Phys. {\bf 43,} 1237\nobreak (2002);\break~{\bf 44},~2724 (2003) (Erratum).}
\pref{\hallg}{R. L. Hall, W. Lucha, and F. F. Sch\"oberl, 
Phys. Lett. A}{320}{127 (2003)}{} 
\bref{\fel}{W. Feller}{An introduction to probability theory and its 
applications, Volume II}{John Wiley, New York, 1971} {Jensen's inequality is 
discussed on p.~153.}

\bref{\as}{M. Abramowitz and I. A. Stegun (eds.)}{Handbook of Mathematical Functions with Formulas, Graphs, and Mathematical Tables}{Dover, New York, 1972}{Modified Bessel functions of the second kind $K_n(z)$ are discussed on pp. 374--377.}
\bref{\ma}{Wolfram Research Inc.}{Mathematica Version 4~~}{1999}{Modified Bessel functions of the second kind $K_n(z)$ are
represented in {\it Mathematica} by $K_n(z)$ = {\tt BesselK(n,z)}.}
\gref{\herb}{I. W. Herbst, Commun. Math. Phys. {\bf 53}, 285 (1977);~{\bf 55},  316 (1977) (addendum).}
\pref{\mart}{A. Martin and S. M. Roy, Phys. Lett. B}{233}{407 (1989)}{}

} 
 
\references{0} 
 
\topskip=20pt 
\trr 
\ppl{CUQM-104}\ppl{HEPHY-PUB 780/04}\ppl{math-ph/0405025} 
\ppl{May 2004}\medskip 
\vskip 0.4 true in 
\centerline{\trbig \ptitle} 
\medskip 
\centerline{\trbig \ptitlee} 
\vskip 0.4true in 
\baselineskip 12 true pt 
\centerline{\bf Richard L.~Hall$^1$, Wolfgang Lucha$^2$, and Franz 
F.~Sch\"oberl$^3$}\medskip 
\nll $^{(1)}${\sl Department of Mathematics and Statistics, Concordia 
University, 1455 de Maisonneuve Boulevard West, Montr\'eal, Qu\'ebec, Canada 
H3G 1M8} 
\nll $^{(2)}${\sl Institut f\"ur Hochenergiephysik, \"Osterreichische 
Akademie der Wissenschaften, Nikolsdorfergasse 18, A-1050 Wien, Austria} 
\nll $^{(3)}${\sl Institut f\"ur Theoretische Physik, Universit\"at Wien, 
Boltzmanngasse 5, A-1090 Wien, Austria} 
  
\nll{\sl rhall@mathstat.concordia.ca, wolfgang.lucha@oeaw.ac.at, 
franz.schoeberl@univie.ac.at} 
\bigskip\medskip 
\baselineskip = 18true pt 
 
\centerline{\bf Abstract}\medskip 
 
\noindent We study the lowest energy $E$ of a relativistic system of $N$ 
identical bosons bound by pair potentials of the form $V(r_{ij}) = g(r_{ij}^2)$ in three 
spatial dimensions. In natural units $\hbar=c=1$ the system has the 
semirelativistic `spinless-Salpeter' Hamiltonian
$$H= \sum_{i=1}^N\sqrt{m^2+\mb{p}_i^2}+\sum_{j>i=1}^N g(|\mb{r}_i-\mb{r}_j|^2),$$
\nl where $g$ is monotone increasing and has convexity $g''\geq 0.$  We use `envelope theory' to derive formulas for
general lower energy bounds and we use a variational method to find complementary upper bounds valid for all $N \geq 2.$ In particular, 
 we determine the energy of the \hi{N}{body} oscillator $g(r^2) = c r^2$ with error less than $0.15\%$ for all $m \geq 0,$ $N\geq2,$ and $c > 0.$
\medskip\noindent PACS: 03.65.Ge, 03.65.Pm, 11.10.St 
\np 

\title{I.~~Introduction} 
We consider a system of $N$ identical bosons interacting by attractive pair potentials 
$V(r_{ij})$ and obeying the semirelativistic spinless Salpeter equation\sref{\bse, \se}. The Hamiltonian governing the dynamics of the \hi{N}{particle} problem is given by
 $$H= \sum_{i=1}^N\sqrt{m^2+\mb{p}_i^2}+\sum_{j>i=1}^N V(|\mb{r}_i-\mb{r}_j|)\eqno{(1.1)}$$
and represents a model system having a relativistically correct expression for the kinetic energy and a static pair potential. One of the reasons for considering such a model is that the extension to the many-particle case poses
no fundamental technical problems beyond what are already present in the one-body problem, namely the square root in the kinetic energy and the non-locality that the definition\sref{\lieb} of the Hamiltonian entails. Our lower bounds use the necessary permutation symmetry of the \hi{N}{boson} problem to effect a `reduction' to an almost equivalent \hi{2}{body} problem\sref{\hallpost,\hallferm}.  The purpose of the present paper is first to use envelope theory\sref{\halla-\halle} to extend our specific energy lower bounds for the harmonic oscillator\sref{\hallf} to apply to smooth transformations of the oscillator having the general form $V(r) = g(r^2),$ where $g$ is monotone increasing and of positive convexity ($g''\geq 0$). Secondly, we show that the earlier upper energy bounds (via a Gaussian trial function) for the oscillator $V(r) = cr^2$ can be considerably sharpened; this improvement is carried over to the larger class of pair potentials.  We have already shown this\sref{\hallg} for the ultrarelativistic case $m = 0$ of the pure oscillator. In this paper we shall generalize these oscillator results to $V(r) = g(r^2)$ and $m \geq 0.$ For the oscillator $V(r) = c r^2$ itself, the new bounds are separated by less than $0.15\%$ for all  $m \geq 0,$ $c > 0,$ and $N\geq 2.$
\par
In Section~II we recall some fundamental formulas concerning the one-body harmonic oscillator with Hamiltonian $\sqrt{m^2 + \mb{p}^2} + r^2$ and lowest energy $e(m).$ This problem does not have an exact analytical solution but can be easily solved numerically to yield $e(m)$ to arbitrary accuracy; this result is necessary for our \hi{N}{body} lower bounds. As distinct from our earlier work\sref{\hallf}, in this paper we eschew the \hi{P}{representation} and its concomitant scaling subtleties, and base all our lower bounds on the function $e(m)$ itself.\medskip
  
In Section~III we turn to the principal topic of this paper, namely potentials which are smooth transformations $V(r) = g(r^2)$ of the oscillator potential.  If $g$ is convex ($g'' \geq 0$), the graph of $V(r)$ lies above `tangential potentials' $V^{(t)}(r)$ with the general form $V^{(t)}(r) = a(t) + b(t) r^2,$ where $t = \hat{r}^2$ is the point of contact with the potential $V(r)$ itself.  As $t>0$ varies, $\{V^{(t)}(r)\}$ represents a family of shifted oscillators.  Envelope theory allows one to construct energy lower bounds based on this fundamental geometrical idea.  In Section~IV we construct variational upper bounds by use of a translation-invariant Gaussian trial function. In Section~V we look at the ultrarelativistic case $m\rightarrow 0,$ and in Section~VI we apply our general results to some examples from the family $V(r) = c r^q,\ q \geq 2.$   
\title{II.~~The one-body oscillator problem} 
We consider the one-body problem with Hamiltonian 
$$H_1=\sqrt{m^2+\mb{p}^2}+r^2\quad\to\quad e(m),\eqno{(2.1)}$$where, for 
coupling $c=1,$ $e(m)$ is the lowest eigenvalue as a function of the 
mass $m.$ In the momentum-space representation, we have an 
equivalent problem with Hamiltonian
$$\tilde H_1=-\Delta+\sqrt{m^2+r^2}\quad\to\quad 
e(m).\eqno{(2.2)}$$
Since this Schr\"odinger problem is easy to solve 
numerically to arbitrary accuracy, we shall take the position that $e(m)$ is 
`known' and at our disposal. We note that in the \hy{large}{m} 
(nonrelativistic or Schr\"odinger) limit, we have
$$e(m)\simeq e_{\rm NR}(m)= 
m+{{3}\over{\sqrt{2m}}}.\eqno{(2.3)}$$ 
The graph of $e(m)-m$ is shown in Figure~1: $e(m)$ 
is monotone increasing with $m;$ $e(m)-m,$ however, is monotone {\it 
decreasing\/}, in agreement, for large $m,$ with the Feynman--Hellmann 
theorem for the corresponding nonrelativistic case.\medskip
It remains now to use scaling to generalize these results.  This is necessary for our later
application to the \hi{N}{body} problem.  For the energy
of a more general one-body problem in 
which the kinetic-energy term is multiplied by the positive factor $\beta,$ 
 the coupling $\gamma>0$ is included, and a further parameter $\lambda > 0$ is allowed for,
 we have, by scaling arguments,
$$H_1=\beta\sqrt{m^2+\lambda\mb{p}^2}+\gamma r^2
\quad\to\quad\varepsilon(m,\beta,\gamma\lambda) = \left({{\beta^2}\gamma\lambda}\right)^{1/3} 
e\left(m\left({\beta\over{\gamma\lambda}}\right)^{1/3}\right).\eqno{(2.4)}$$
\np 
\title{III.~~Energy lower bound for $V(r) = g(r^2)$ by envelope theory} 
Our hypothesis is that $V(r) = g(r^2),$ where the smooth transformation function $g$ is monotone increasing and its convexity is positive or zero.  That is to say, we shall assume $g'' \geq 0.$ These assumptions imply a relation between $V(r)$ and a `tangential potential' $V^{(t)}(r)$ given explicitly by
$$V(r) \geq V^{(t)}(r) = g(t) - tg'(t) + g'(t)r^2 = a(t) + b(t)r^2,\eqno{(3.1)}$$
\nl where $t = \hat{r}^2$ is the point of contact between the tangential potential and the potential.  For each fixed $t,$ the tangential potential has the form $a + br^2$ of a shifted oscillator.
 This potential inequality  induces, in turn, a spectral inequality as an immediate consequence of the min-max characterization of the spectrum of the Hamiltonian. It is the task of envelope theory\sref{\halla-\halle} to generate expressions for this spectral inequality.\medskip

The kinetic-energy term in the Hamiltonian $H$ does not have the kinetic energy of the center-of-mass removed. Thus the wave function we use must satisfy two fundamental symmetries: translation invariance and boson permutation symmetry (in the individual-particle coordinates). 
Jacobi relative coordinates may be defined with the aid of an orthogonal 
matrix $B$ relating the column vectors of the new $[\rho_i]$ and old 
$[\mb{r}_i]$ coordinates given by $[\rho_i]=B[\mb{r}_i].$ The 
first row of $B$ defines a center-of-mass variable $\rho_1$ with every entry 
$1/\sqrt{N},$ the second row defines a pair distance $\rho_2 
=(\mb{r}_1-\mb{r}_2)/\sqrt{2},$ and the $k\!$th row, $k\ge 2,$ has the first 
$k-1$ entries $B_{ki}=1/\sqrt{k(k-1)},$ the $k\!$th entry 
$B_{kk}=-\sqrt{(k-1)/k},$ and the remaining entries zero. We define the 
corresponding momentum variables by $[\pi_i]=(B^{-1})^{\rm 
t}[\mb{p}_i]=B[\mb{p}_i].$ Let us suppose that the (unknown) exact normalized boson ground-state wave function 
for the \hi{N}{body} harmonic-oscillator problem with $V(r) = c r^2$ is $\Psi = \Psi(\rho_2, \rho_3,\dots,\rho_N)$ corresponding to energy $E.$  Boson symmetry is a powerful constraint that greatly reduces the complexity of this problem.  We immediately obtain\sref{\hallf, Eq. (2.3)} the `reduction'
$$E = (\Psi, H\Psi) = \left(\Psi, \left[N\sqrt{m^2+\mb{p}_N^2}+ {{N(N-1)}\over 2} c |\mb{r}_1-\mb{r}_2|^2\right]\Psi\right).\eqno{(3.2)}$$
\nl Since $|\mb{r}_1-\mb{r}_2|^2 =  2\rho_2^2,$ in terms of the Jacobi relative coordinates this becomes 
$$E = \left(\Psi, \left[N\sqrt{m^2 + \left({{\pi_1}\over{\sqrt{N}}}- \sqrt{{N-1}\over{N}}\pi_N\right)^2} + N(N-1)c   \rho_2^2\right]\Psi\right).\eqno{(3.3)}$$
\nl The lemma proved in\sref{\hallf} allows us to remove the term in the center-of-mass momentum operator $\pi_1$ from inside the square root. Boson permutation symmetry furthermore implies\sref{\hallf, Eq. (2.5)}
$$(\Psi,\rho_2^2 \Psi) = (\Psi,\rho_N^2 \Psi),\eqno{(3.4)}$$ 
even though the wave function $\Psi$ may not be symmetric in the relative coordinates.  These results lead to the final reduction
$$E = \left(\Psi, \left[N\sqrt{m^2 + {{N-1}\over{N}}\pi^2_N} + N(N-1)c \rho_N^2\right]\Psi\right).\eqno{(3.5)}$$
\nl If we now write $\mb{r} = \rho_N$ and $\mb{p} = \pi_N,$ we see that the exact energy $E$ can be written in the form $E = (\Psi, {\cal H}\Psi),$ in which ${\cal H}$ is the Hamiltonian for a one-body problem given by
$${\cal H} = \beta\sqrt{m^2+\lambda\mb{p}^2}+ \gamma c\mb{r}^2,\eqno{(3.6)}$$
\nl with
$$\beta = N,\quad \lambda = {{N-1}\over{N}},\quad {\rm and}\quad \gamma =  N(N-1).$$
\nl It follows that the exact energy $E$ of the oscillator system is bounded below by ${\cal E},$ the bottom of the spectrum of the one-body Hamiltonian ${\cal H}.$

\nl Thus, for the harmonic oscillator itself, we have from (3.5) and (2.4)
\nll{\bf Theorem~1}
\nll{\it 
A lower bound to the ground-state energy eigenvalue $E$ of the semirelativistic 
\hi{N}{body} Hamiltonian 
$$H= \sum_{i=1}^N\sqrt{m^2+\mb{p}_i^2}+\sum_{j>i=1}^N c |\mb{r}_i-\mb{r}_j|^2,\quad c > 0,\eqno{(3.7)}$$
\nl is provided by the formula

$$E\geq\left({{\beta^2}\gamma c \lambda}\right)^{1/3} 
e\left(m\left({\beta\over{\gamma c \lambda}}\right)^{1/3}\right),\eqno{(3.8)}$$
\nl where 
$$\beta = N,\quad \lambda = {{N-1}\over N},\quad \gamma = N(N-1).$$
}

\nl This lower bound yields the exact energy in the Schr\"odinger limit $m\rightarrow \infty.$
  If we consider the potential $V(r) = g(r^2)$ and use the potential lower bound (3.1), we can maximize the resulting lower bound provided by Theorem~1 to obtain
\goodbreak
\nll{\bf Theorem~2}
\nll{\it 
A lower bound to the ground-state energy eigenvalue $E$ of the semirelativistic 
\hi{N}{body} Hamiltonian 
$$H= \sum_{i=1}^N\sqrt{m^2+\mb{p}_i^2}+\sum_{j>i=1}^N g(|\mb{r}_i-\mb{r}_j|^2),\quad g'>0,\quad g''\geq 0,\eqno{(3.9)}$$
\nl is provided by the formula
$$E \geq \max_{t > 0}\left[m\beta{{e(\nu)}\over{\nu}} 
 + {{\gamma}\over 2} (g(t)-tg'(t))\right],\eqno{(3.10)}$$
\nl where 
$$\beta = N,\quad \lambda = {{N-1}\over N},\quad \gamma = N(N-1),\quad \nu = m \left({{\beta}\over{\gamma \lambda g'(t)}}\right)^{1/3}.$$
}
\par If we consider the family of pure-power potentials of the form $V(r) = cr^q,$ then for the harmonic oscillator $ q = 2,$ we use Theorem~1; for more general potentials, with $q > 2,$ we have $V(r) = g(r^2) = g(t) = c t^{q/2}.$ Consequently, we must in this case make the explicit substitutions: 
$$a(t) = g(t)-tg'(t) = -c \left({q\over 2} - 1\right) t^{q/2}\quad{\rm and}\quad b(t) = g'(t) = {{cq}\over 2}t^{(q-2)/2}.\eqno{(3.11)}$$ 
\title{IV.~~Variational upper bounds} 
Improvement over the previous upper energy bounds\sref{\hallf} for the oscillator will be obtained in this paper by avoiding the loosening incurred by use of Jensen's inequality\sref{\lieb,\fel}.  This goal has already been achieved\sref{\hallg} for the ultrarelativistic special case $m = 0$ of the \hi{N}{body} harmonic-oscillator problem. We shall now extend this to more general problems with attractive potential $V(r)$ and $m\geq 0.$ \medskip
We use a Gaussian wave function of the form 
$$\Phi(\rho_2,\rho_3,\dots,\rho_N)= 
C\exp\left(-{\alpha\over 2}\sum_{i=2}^N\rho_i^2\right),\quad\alpha>0,\eqno{(4.1)}$$
where 
$C$ is a normalization constant. The factoring property of this 
function, the boson-symmetry reduction leading to (3.5), and the additional fact that $\Phi$ is also symmetric under exchange of the {\it relative} coordinates allows us to write 
$\mb{r} = \rho_2,$ and $\mb{p} = \pi_N\rightarrow \pi_2,$ and finally
$$E\le 
\beta\left(\phi,\sqrt{m^2+\lambda\mb{p}^2}\ \phi\right) 
+{\gamma\over 2}\left(\phi,V(\sqrt{2}r)\phi\right),\eqno{(4.2)}$$
\nl where
$$\beta = N,\quad \lambda = {{N-1}\over{N}},\quad \gamma = N(N-1),$$
\nl and the function $\phi(r)$ is given by
$$\phi(r)=\left({{\alpha}\over{\pi}}\right)^{3/4}\exp\left(-{{\alpha 
r^2}\over 2}\right).\eqno{(4.3)}$$

The kinetic-energy integral may be written in terms of modified Bessel functions of the second kind\sref{\as,\ma}, which we now discuss.  The calculation is best carried out in momentum space with the aid  of the three-dimensional Fourier transform ${\cal F}$.  We have
$$\phi(r) \buildrel{\cal F}\over{\longrightarrow} \tilde{\phi}(k) = \left({{1}\over{\alpha\pi}}\right)^{3/4}\exp\left(-{{
k^2}\over {2\alpha}}\right).\eqno{(4.4)}$$
\nl Thus the expectation of the kinetic energy becomes

$$\langle K\rangle = \beta\left(\tilde{\phi},\sqrt{m^2+\lambda k^2}\ \tilde{\phi}\right) = {{4\pi\beta}\over{(\alpha\pi)^{3/2}}}\int_0^{\infty}\exp\left(-{{k^2}\over{\alpha}}\right)\sqrt{m^2+\lambda k^2}\  k^2 dk.\eqno{(4.5)}$$   
\nl We may write this integral in the form
$$\langle K\rangle = {{\beta m\mu}\over{\sqrt{2\pi}}}\exp\left({{\mu^2}\over{4}}\right)K_1\left({{\mu^2}\over{4}}\right),\quad \mu = m\left({2N\over{(N-1)\alpha}}\right)^{1/2},\eqno{(4.6)}$$
\nl where $K_{\nu}(z)$ is a modified Bessel function of the second kind\sref{\as,\ma}.\medskip
The potential-energy integral will depend on the choice of $V(r).$ For the family $V(r) = c\ \sgn(q)r^q,$  which we shall study in Section~V, the integrals may be expressed in terms of the gamma function. Explicitly we have 
$$\langle V\rangle = \left(\phi, \left({{c\ \sgn(q)\gamma}\over{2}}\left(r\sqrt{2}\right)^q\right)\phi\right) = {{c\  \sgn(q)\gamma}\over{\sqrt{\pi}}}\Gamma\left({{3+q}\over{2}}\right)\left({{\mu\sqrt{\lambda}}\over{m}}\right)^q.\eqno{(4.7)}$$
\nl With the results in this form we can use the parameter $\mu$ as a variational parameter.  We have therefore established
\nll{\bf Theorem~3}
\nll{\it 
 For fixed $m > 0,$ $q > -1,$ $c > 0,$ $N \geq 2,$ $\beta = N,$ $\gamma = N(N-1),$ and $\lambda = (N-1)/N,$ the lowest energy $E$ of the
\hi{N}{boson} problem for the pair potential $V(r) = c\ \sgn(q)r^q$ is given by the inequality
$$E\  \leq\  \min_{\mu > 0}\left[{{\beta m\mu}\over{\sqrt{2\pi}}}\exp\left({{\mu^2}\over{4}}\right)K_1\left({{\mu^2}\over{4}}\right) + 
{{c\ \sgn(q)\gamma}\over{\sqrt{\pi}}}\Gamma\left({{3+q}\over{2}}\right)\left({{\mu\sqrt{\lambda}}\over{m}}\right)^q\right].
\eqno{(4.8)}$$
}

\nl We have allowed $q > -1$ here since the upper bound easily accommodates this family of potentials in three spatial dimensions. For $q < -1,$ there is no discrete spectrum. In the gravitational case $q = -1$ the minimum upper bound exists provided the coupling is not too large: specifically we require
$${{c\gamma}\over{4\beta}}\sqrt{{2\over{\lambda}}} = {c\over 2}\sqrt{{{N(N-1)}\over 2}} < 1.\eqno{(4.9)}$$
\nl  This situation is of course well known from the two-particle attractive Coulomb problem\sref{\herb,\mart}. At present we only have complementary lower bounds for $q \geq 2.$
\title{V.~~The ultrarelativistic limit}
The ultrarelativistic case $m\rightarrow 0$ may be obtained from Theorems~2 and 3 as a special case. The Hamiltonian for this problem is given explicitly by
$$H= \sum_{i=1}^N \sqrt{\mb{p}_i^2}+\sum_{j>i=1}^N c |\mb{r}_i-\mb{r}_j|^q,\quad c > 0,\ q \geq 2.\eqno{(5.1)}$$
\nl For the lower bound we use $g(t) = c t^{q/2}.$ The upper bound may either be treated separately or taken from  (4.8) by means of the limit $\lim_{z \rightarrow 0}zK_1(z) = 1.$  The bounds we obtain are given by
$$ C\left[{{z_0}\over 3}\right]^{{3q}\over{2(1+q)}}\ \leq\ E\ \leq\ {C\over{\sqrt{\pi}}}
\left[2\Gamma\left({{3+q}\over{2}}\right)\right]^{1\over{1+q}},\eqno{(5.2)}$$
\nl where $z_0 \approx 2.33810741$ is the first zero of the Airy function $\Ai(z),$ satisfying $\Ai(z_0) = 0,$ and the common factor $C$ is given by
$$C = \left({{cq}\over{2}}\right)^{1\over{1+q}}
\left(1+{1\over q}\right)\left(N(N-1)\right)^{{2+q}\over{2(1+q)}}2^{{3q}\over{2(1+q)}}.$$
\nl As $m$ increases from zero, the power-law bounds become closer monotonically with $m;$ thus the $m = 0$ case provides an upper bound to the error for all $m.$  Since we have explicit formulas for the bounds in terms of $N,$ we are able to make definite statements concerning the percentage separations of the bounds for all $N.$ If we take the energy estimate $\bar{E}$ to be the average of the bounds, the exact energy $E$ is determined by $\bar{E}$ to within $0.15\%$ for the harmonic oscillator $q = 2,$ and to $3.6\%$ for the cubic `oscillator' $q = 3.$ In the nonrelativistic limit $m\rightarrow \infty$ the harmonic-oscillator bounds $q = 2$ coalesce to the exact solution of the Schr\"odinger \hi{N}{body} problem\sref{\hallf}.
\title{VI.~~Examples}
The examples we consider are from the family $V(r) = c r^q.$  In order to have a lower bound, we restrict the power to $q \geq 2.$  We revisit the oscillator problem because we have considerably improved the upper bound since Ref.\sref{\hallf}.  Graphs of the lower bounds alone are shown in Figure~2.  The percentage separations are bounded above by the separations at $m = 0,$ which are there less than $0.15\%.$ With the notation $E_N(m)$ we have explicitly, for the oscillator $q = 2,$
 that the lower and upper estimates have numerical values $E_8^{\rm L}(1) = 35.86383$ and $E_8^{\rm U}(1) = 35.89953,$ respectively.  Thus the average of these values determines $E_8(1)$ in this case with error less than $0.05\%.$  As we leave the oscillator and increase $q$ beyond $q = 2,$  the bounds become less sharp.  For $q = {5\over 2}$ we show the corresponding bounds in Figure~3: here the bounds are separated for all $m$ by less than $1.43\%.$ The corresponding graphs for the cubic `oscillator' $q = 3$ are shown in Figure~4;  in this case the maximum percentage separation (again for all $N\geq 2$ and $m\geq 0$) is $3.6\%.$  
\title{VII.~~Conclusion}
The necessary permutation symmetry of the states of a system of identical particles is a powerful constraint.  The approximate `reduction' of the \hi{N}{body} problem to a scaled two-body problem is most striking for systems of bosons, or for systems which are compatible with the assumption of permutation symmetry in the spatial variables alone\sref{\hallpost}.  For systems of fermions, the reduction is to a sum over two-body energies\sref{\hallferm}.  For the Schr\"odinger harmonic-oscillator problem, the boson reduction is complete in the sense that the \hi{N}{body} energy is given exactly by the energy of a two-body problem.  A lower bound by this type of reduction is possible provided either the kinetic-energy term or the potential-energy term has a quadratic form: this allows us to replace, for example, the `mixed' pair $\{\pi_N,\ \rho_2\}$ by $\{\pi_N,\ \rho_N\}$ in the reduced two-body Hamiltonian ${\cal H}.$  For the Salpeter problem discussed in this paper, a quadratic form is present in the potential term of the oscillator, and the lower bound obtained for this base problem is then applicable to other problems whose potentials $V(r)$ have the form of smooth convex transformations $g(r^2)$ of the oscillator.  The extension beyond the oscillator is effected by the use of `envelope theory'.  A reduction is also used for our upper bound, but this reduction is allowed for general potentials and for a different reason. The trial function must be a translation-invariant boson function; but we have chosen a Gaussian trial function which has an additional symmetry, namely, it is also symmetric in the relative coordinates.  It is this latter symmetry which completes the reduction in the case of the upper bound.  Because of all these symmetries, what starts out as a complex many-body problem, appears in the end, for the purpose of finding energy bounds, as a one-body problem.        
\title{Acknowledgements} 
 
 Partial financial support of this work under Grant No. GP3438 from the Natural 
Sciences and Engineering Research Council of Canada, and the hospitality of the 
Institute for High Energy Physics of the Austrian Academy of Sciences in Vienna,
 is gratefully acknowledged by one of us [RLH].\medskip 
\np 
\references{1} 
\hbox{\vbox{\psfig{figure=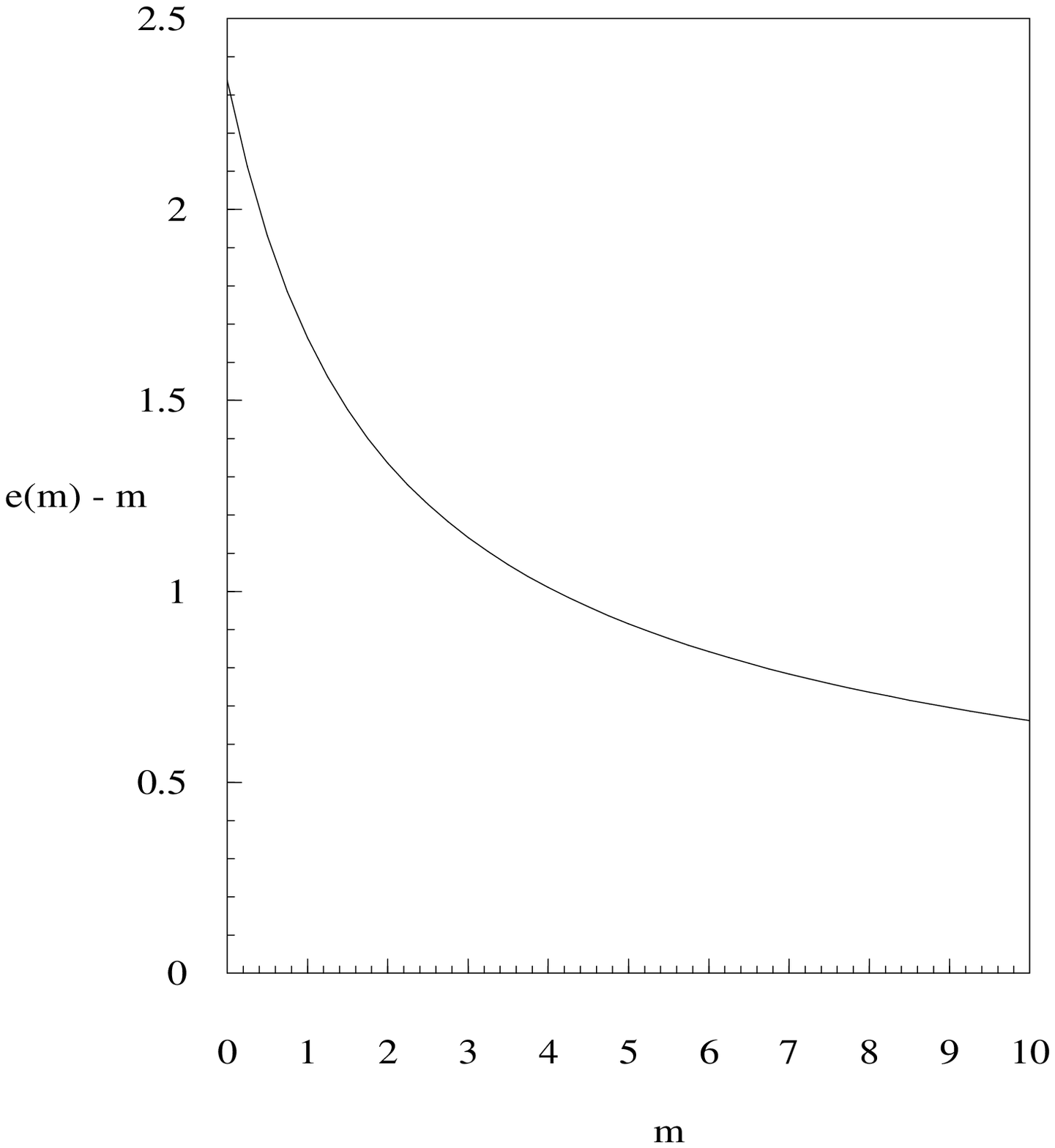,height=6in,width=5.4in,silent=}}}
\nl{\bf Figure 1.}~~The energy function $e(m) - m$ of the one-body problem 
defined~by (2.1). 
 
\hbox{\vbox{\psfig{figure=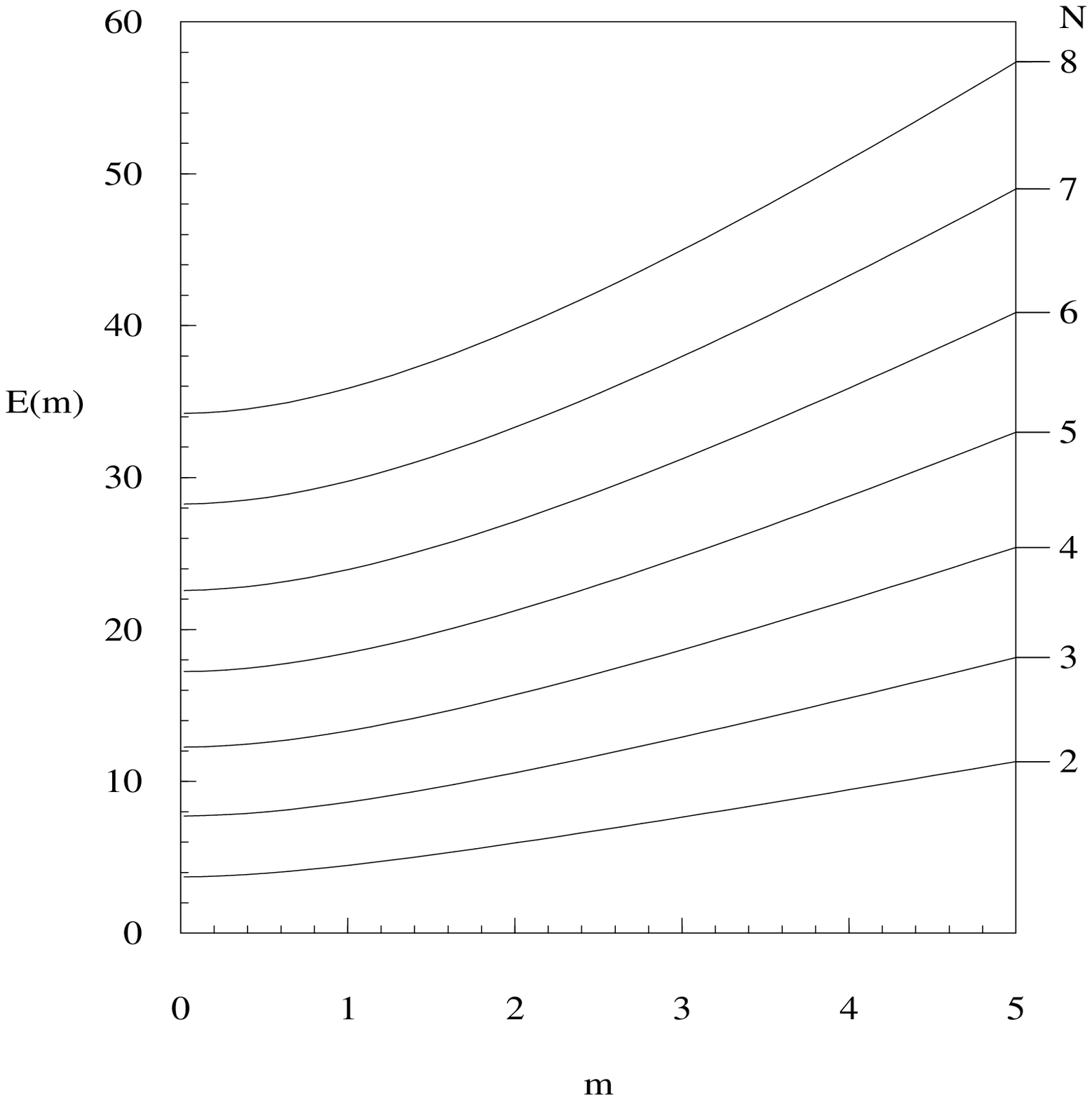,height=6in,width=5in,silent=}}}
\nl{\bf Figure 2.}~~The ground-state energy $E(m)$ of the relativistic \hi{N}{boson} harmonic-oscillator problem $V(r) = r^2$ 
for $N=2,3,\dots,8.$ The figure shows the lower bounds given by Eq.~(3.8): the upper bounds are everywhere less than $0.15\%$ above these curves and are indistinguishable on the graph.  In the Schr\"odinger limit $m\rightarrow\infty$ the upper and lower bounds coalesce to the exact energies.

\np\hbox{\vbox{\psfig{figure=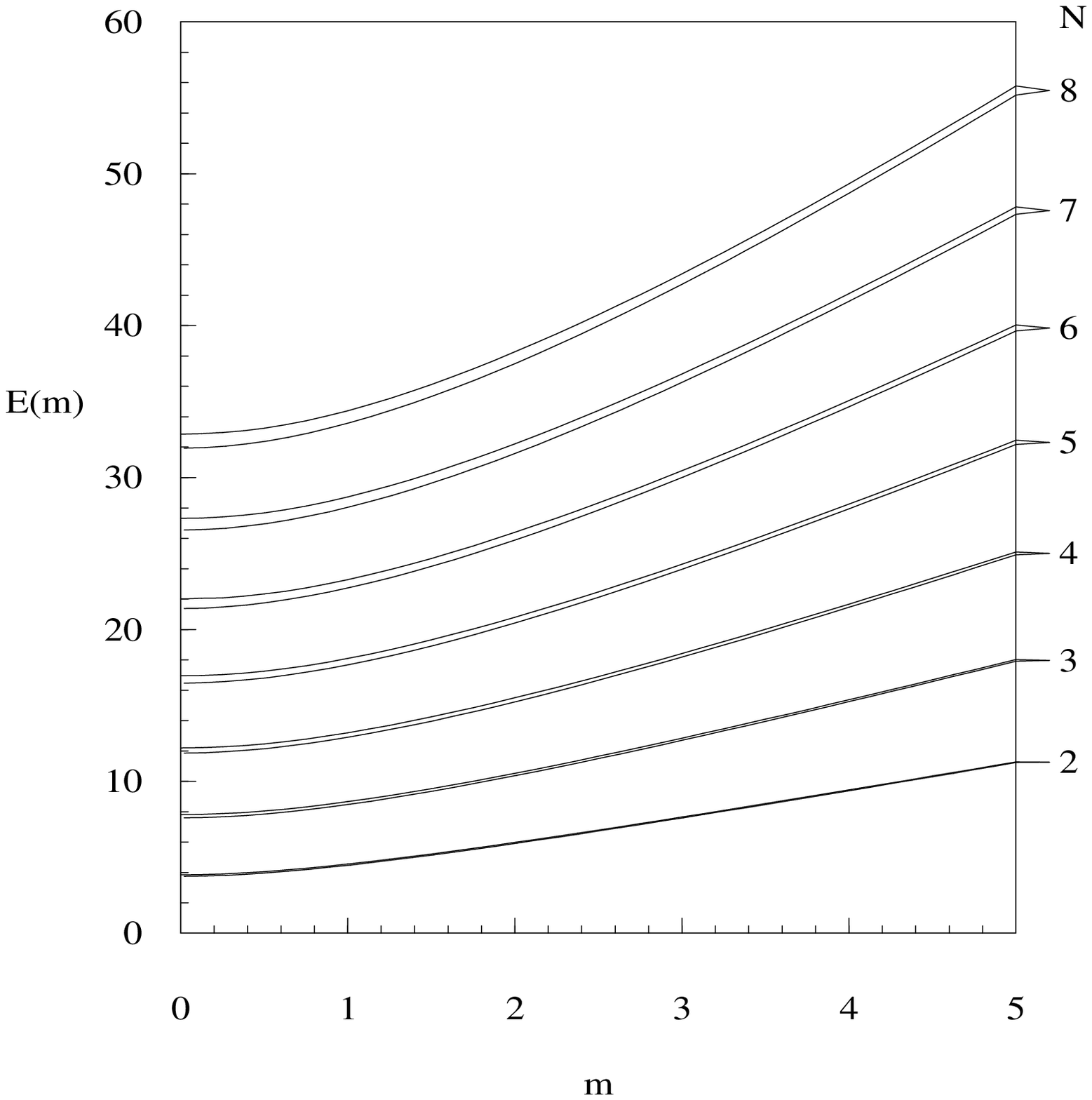,height=6in,width=5in,silent=}}}
\nl{\bf Figure 3.}~~Upper and lower energy bounds for the ground-state energy $E(m)$ of the relativistic \hi{N}{boson} problem corresponding to $V(r) = r^{5/2}$ for $N=2,3,\dots,8.$ The percentage errors are maximum for $m = 0$ where they determine the energies (for all $N)$ with error less than $1.43\%.$ 

\np\hbox{\vbox{\psfig{figure=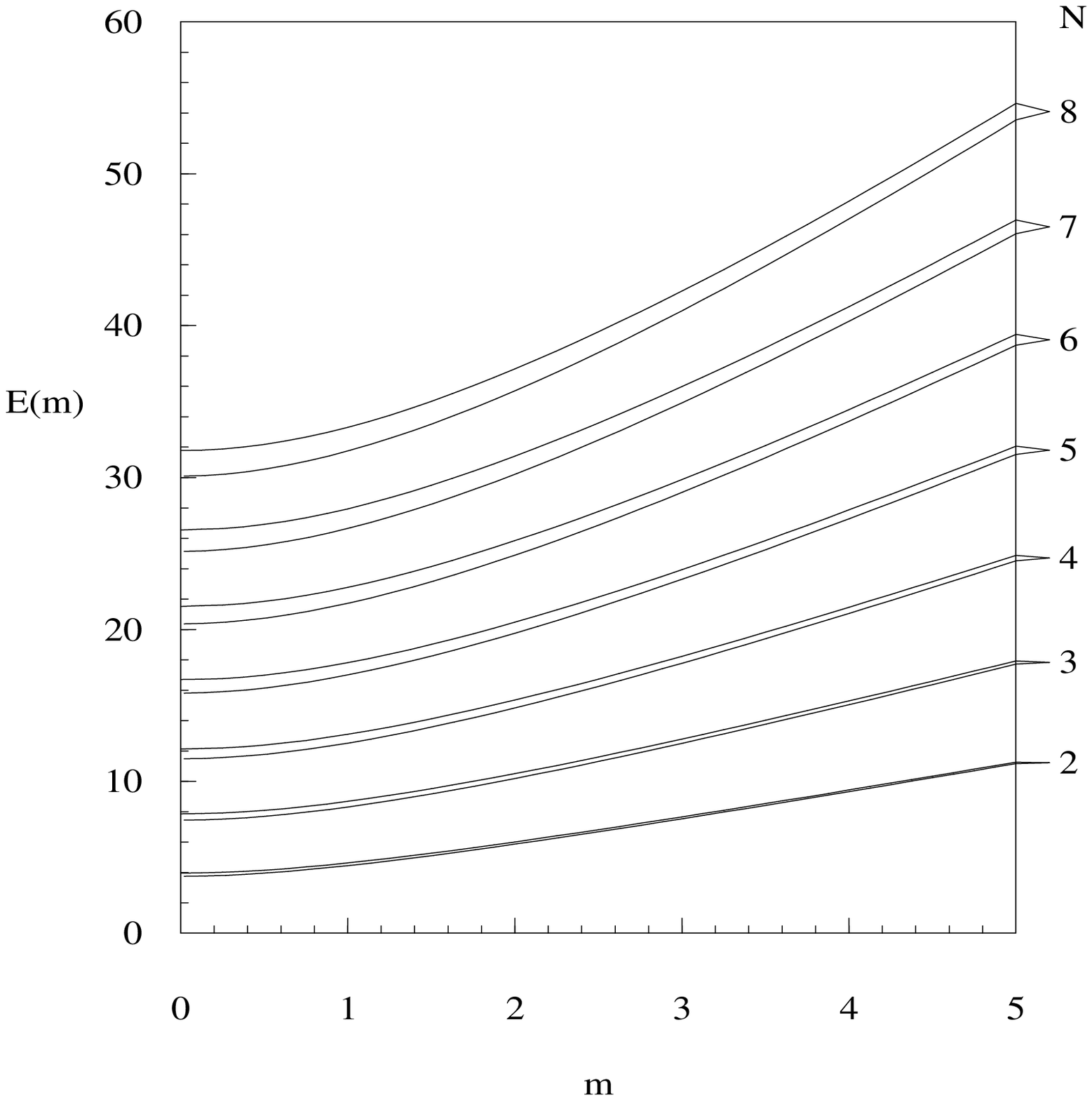,height=6in,width=5in,silent=}}}
\nl{\bf Figure 4.}~~Upper and lower energy bounds for the ground-state energy $E(m)$ of the relativistic \hi{N}{boson} problem corresponding to $V(r) = r^{3}$ for $N=2,3,\dots,8.$ The percentage errors are maximum for $m = 0$ where they determine the energies (for all $N)$ with error less than $3.6\%.$

\end